# The Surface response around a sharply resonant surface polariton mode is simply a Lorentzian

**J.C. DE AQUINO CARVALHO[1,2,A] AND D. BLOCH[1]\***

[1] *Laboratoire de physique des lasers, UMR7538 du CNRS and Université Sorbonne Paris Nord, 99 Av. JB Clément, F-93430 Villetaneuse*
[2] *Departamento de Física, Universidade Federal de Pernambuco, Recife, PE 50670-901, Brazil*
*Corresponding author: daniel.bloch@univ-paris13.fr*

A : https://orcid.org/0000-0003-4521-8807

**At the planar interface between a material and vacuum, the complex surface response $S(\omega)=[\varepsilon(\omega)-1]/[\varepsilon(\omega)+1]$, with $\varepsilon(\omega)$ the relative complex dielectric permittivity of the material, exhibit resonances, typical of the surface polariton modes, when $\varepsilon(\omega) \sim -1$. We show that for a moderately sharp resonance, $S(\omega)$ is satisfactorily described with a mere (complex) Lorentzian, independently of the details affecting the various bulk resonances describing $\varepsilon(\omega)$. Remarkably, this implies a quantitative correlation between the resonant behaviors of $\mathfrak{Re}[S(\omega)]$ and $\mathfrak{Im}[S(\omega)]$, respectively associated to dispersive and dissipative effects in the surface near-field. We show that this "strong resonance" approximation easily applies, and discuss its limits, based upon published data for sapphire, $CaF_2$ and $BaF_2$. Extension to interfaces between two media or to a non planar interface is briefly considered.**

Surface polaritons are electromagnetic field modes constrained by boundary conditions at an interface [1], whose amplitude decay exponentially away from the interface (evanescent field). Such modes occur for specific conditions, dependent on the properties of the materials at the interface. For an elementary planar interface between a homogeneous (non magnetic) material, characterized by a frequency-dependent relative permittivity $\varepsilon_1(\omega) = \varepsilon_1'(\omega) + i\,\varepsilon_1''(\omega)$, and a transparent dielectric medium ($\varepsilon_2$) or vacuum (*i.e.* $\varepsilon_2 = 1$), these surface modes appear only when $\varepsilon_1'(\omega) < 0$. The spatial decay of the surface polariton from the interface, and its limited propagation length along the interface, owing to losses, are governed by the spectral features of $\varepsilon_1(\omega)$. For a vacuum interface, the optimal confinement of the polariton around the interface occurs when $\varepsilon_1(\omega) + 1 \sim 0$, *i.e.* $\varepsilon_1'(\omega) \sim -1$, while $\varepsilon_1''(\omega)$ is the key parameter to evaluate the depth of this confinement, and the limited propagation along the interface due to losses.

Numerous techniques have been developed to evidence experimental effects induced by the presence of surface polaritons, in the case of an elementary planar interface, or for more complex systems, *e.g.* involving structured or layered interfaces, or nanoparticles. Predicting the spectral properties of the surface polariton requires the knowledge of the complex spectrum of the relative permittivity $\varepsilon(\omega)$, or equivalently, of the complex optical index. This is hardly obtained through direct experimental measurements, as optical regions of transparency [$\varepsilon''(\omega) = 0$] have to be covered, along with spectral domains characterized by a remarkably strong absorption [2]. Aside from ellipsometry [3], whose spectral range is limited, experimental techniques currently rely on the spectral recording of measurement a single real quantity, such as reflectance.

The complex behavior of $\varepsilon(\omega)$ is hence extrapolated, *e.g.* through the Kramers-Kronig causality relationship, or by "guessing" an analytical expression for $\varepsilon(\omega)$, whose coefficients are numerically fitted, with respect to the experimental data, often acquired over a limited spectral domain. If classical but limited descriptions for metals are given by the Drude or the plasma model, more specific models are used for dielectric materials, based upon the summing-up of elementary *bulk* resonances. Each bulk resonance, associated to a vibration in the volume, has to be described appropriately, possibly with an elementary three- or four-parameter model or with a more complex one —notably accounting for the thermal population of phonons. Alternately, and often for differing spectral ranges, one may directly measure the optical constants ($n$, $\kappa$) [with $n$ the refractive index, and $\kappa$ the extinction coefficient, *i.e.* the absorption coefficient over $\lambda/4\pi$, $\lambda$ being the optical wavelength in vacuum]. Tabulated values of optical constants [2] are often derived from a variety of methods, including for some spectral ranges the evaluation of the dielectric permittivity. Indeed, through the general equation relating permittivity and optical constants:

$$\varepsilon(\omega) = [n(\omega) + i\,\kappa(\omega)]^2 \qquad (1)$$

relevant information concerning surface polaritons can also be obtained from the optical constants.

Here we show that the direct use of these optical constants ($n$, $\kappa$) is efficient for the prediction, around the surface polariton resonance, of the (complex) surface response:

$$S(\omega) = [\varepsilon(\omega)-1]/[\varepsilon(\omega)+1] \qquad (2)$$

This complex surface response is particularly relevant to determine resonant atom-surface coupling [4,5], with $\Re[S(\omega)]$ associated to dispersion forces and $\Im[S(\omega)]$ to a real energy transfer. It also appears when describing the near-field properties of thermal emission [6], notably the local density of states, or for the near-field limit of the Casimir interaction [7,8], a macroscopic attraction induced by quantum fluctuations, which takes place between two neighboring neutral media.

As long as the surface polariton exhibits a reasonably sharp resonance, we show, on a general basis, that a limited information on the optical constants $(n, \kappa)$ around the surface resonance, suffices to predict in detail the shape of the resonance for $S(\omega)$. $S(\omega)$ merely reduces to a complex Lorentzian, independently of the model required to describe bulk resonances, associated to $\varepsilon(\omega)$. At last, we compare, for well-known materials, our simplified predictions to those derived from elaborate models, justifying that the assumption of a "sharp" surface resonance, needed for our calculation, is not too restrictive.

As discussed in [4], where some key features (width and height of resonances) of surface resonances were tabulated for nearly a hundred of materials —essentially those tabulated in [2] —, sizeable surface resonances essentially occur for frequencies satisfying simultaneously $\kappa(\omega) \sim 1$ and $n(\omega) \ll 1$ (and intrinsically $n > 0$). To characterize the shape of $S(\omega)$ around the resonance, the key point is to note that the surface resonance is strongly shifted with respect to the bulk resonances [4, 5]. This justifies linearizing the frequency variations of the optical coefficients, which evolve smoothly across the surface resonance. Moreover, because $n$ should reach "small" values ($n \ll 1$), it should be close to a minimum. We first define an approximate surface polariton frequency $\omega_p$, stating that $\kappa(\omega_p) = 1$. One then approximates, around $\omega_p$:

$$\kappa(\omega) = 1 + K' \Delta \quad (3)$$

with $\Delta = \omega - \omega_p$, and $K' = \left.\frac{d\kappa}{d\omega}\right|_{\omega_p}$

For $n(\omega)$, it is sufficient to retain, around resonance, that $n(\omega) = n(\omega_p) = n_0$, so that the variations of $n(\omega)$ around $n_0$, which is close to a minimum of $n(\omega)$, can be neglected. Although this may seem a crude approximation, this will appear fully legitimate for the practical examples that we will consider. Indeed, even assuming for consistency a linearized expansion similar to eq.(3), $n(\omega) = n_0 + N' \Delta$, the assumption $n(\omega_p) \ll 1$, along with the intrinsic condition $n(\omega) > 0$, would naturally lead to $|N'| \ll |K'|$.

We start with the surface response expressed with optical index and extinction coefficients:

$$S(\omega) = \frac{(n^2 - \kappa^2 - 1) + 2in\kappa}{(n^2 - \kappa^2 + 1) + 2in\kappa} = 1 - 2\frac{(n^2 - \kappa^2 + 1) - i2n\kappa}{\left(n^2 - \kappa^2 + 1\right)^2 + 4n^2\kappa^2} \quad (4)$$

Hence, one goes to approximations applicable in the vicinity of the surface resonance, characterized by $\kappa \sim 1$, naturally leading to $K'\Delta \ll 1$. This allows to approximate $2n\kappa$ by $2n_0$, and $(n^2 - \kappa^2 + 1)$ by $(n_0^2 - 2K'\Delta)$. In this last case, neglecting $(K'\Delta)^2$ around $\Delta \sim n_0^2/2K'$ may appear problematic at first sight,. Actually, the neglected term, on the order of $(n_0^2/2)^2$, is small with respect to the second term in the denominator $4n^2\kappa^2$ ($\sim 4n_0^2$), owing to $n_0 \ll 1$. This reduces $S(\omega)$ to:

$$S(\omega) = 1 - 2\frac{(n_0^2 - 2K'\Delta) - i2n_0}{\left(n_0^2 - 2K'\Delta\right)^2 + 4n_0^2} = 1 - \frac{2}{\left(n_0^2 - 2K'\Delta\right) + 2in_0} \quad (5)$$

Equation (5) has simply turned the surface response into a complex Lorentzian, whose half-width is $n_0/K'$, amplitude $1/n_0$, and whose centre is shifted from $\omega_p$ by an amount $+ (n_0^2/2K')$. Under the assumption $n_0 \ll 1$, this shift itself remains a small fraction of the half-width surface resonance.

Two remarkable points must be noticed in this elementary result. First, this Lorentzian approximation is obtained independently of the model chosen for the bulk resonances: whatever the complexity of this model, its essential features occur on frequency ranges away from the surface resonance. Second, under the "strong resonance approximation" leading to eq.(5), $\Re[S(\omega)]$ and $\Im[S(\omega)]$, which correspond to different physical effects, usually exhibiting unrelated experimental signatures, appear profoundly coupled around the resonance. This brings strong constraints, as soon as $\Re[S(\omega)]$ -or $\Im[S(\omega)]$- is known at two different frequencies, or if the resonance center can be determined accurately. As an illustration, one may refer to the case of AlSb, where it was shown [4] that the differing values found in the literature for optical constants [2], lead to a major uncertainty on the frequency of the surface resonance. It is here made clear that the uncertainties affecting $\Re[S(\omega)]$ and $\Im[S(\omega)]$ are not independent, and rather are strongly coupled when the resonance is sufficiently sharp.

We now compare our approximate Lorentzian evaluation for $S(\omega)$, with an evaluation derived from analytical expressions for $\varepsilon(\omega)$, as given in the literature for several materials of interest [3,9-13]. This comparison is illustrated in fig.1, which considers sapphire ($Al_2O_3$), $CaF_2$, and $BaF_2$, with published data for these materials not limited to room temperature, but including evaluations of the temperature broadening of the bulk resonances [16], so that $n_0(\omega_p)$ increases with temperature, and can depart from the low values assumed in our calculation.

Among currently available materials, sapphire [9-11] is known to exhibit, along with SiC [6, 14]), very sharp bulk resonances, and also a high quality-factor for its main surface resonance. However, it exhibits complex resonances, with several major modes (2 or 4, depending on the orientation of the birefringent axis) [15]. Despite this complexity of the bulk modes, which induces a complex spectrum for $(n, \kappa)$ —see the top line of fig.1—, the surface resonance is very well described by the approximation with a complex Lorentzian, both for $\Re S$ and $\Im S$ (see fig. 1). Even if the resonances for the bulk values $(n, \kappa)$ undergo a slight broadening [9,11] for T = 1000 K, as compared to T = 300 K, this leads to a slight increase in the minimal value of $n(\omega_p)$ but does not affect the validity of the Lorentzian approximation. The calculated differences $\Delta\Re S$ and $\Delta\Im S$, between the Lorentzian approximation and the actual value calculated with the $\varepsilon(\omega)$ model, allow to visualize (see magnification in fig.1) the high quality of the Lorentzian approximation.

For $CaF_2$ [9,12,13], the resonance is broader than for sapphire, and with a lesser amplitude, but a very satisfactory prediction around the peak of the surface polariton resonance is still obtained. This can be understood by the fact that $n$ remains rather flat, around the Lorentzian resonance. Even at high temperature (T=800 K), $n(\omega_p)$ increases but its value ($\sim 0.3$) remains relatively small: the discrepancies introduced by the Lorentzian approximation are clearly visible on the wings, but remain rather minor.

At T = 300 K, $BaF_2$ exhibits a behavior very similar to the one found for $CaF_2$, but for a resonance at a lower frequency [9,12,13]. The surface resonance itself, analogous to the one found for $CaF_2$, is even slightly more pronounced than for $CaF_2$, illustrating again the essential importance of the value $n_0$, with here $n_0 \sim 0.15$ [13] (NB: according to [2], one would rather have $n_0 \sim 0.25$, see discussion in [13]). Nevertheless, with the lower frequency of the vibration modes, temperature effects are more dramatic for $BaF_2$ than for $CaF_2$. For

T = 800 K, our derivation for the Lorentzian approximation, requiring $n_0 \ll 1$, no longer holds with $n_0 \sim 0.55$. Although the approximated Lorentzian surface response is not dramatically bad, notably for the width and the amplitude reduction, it provides a response shifted by a non negligible fraction of the width. Finally, it is also worth remarking that, on the blue side of these broad $CaF_2$ or $BaF_2$ resonances, one has $n \geq \kappa$, implying that $\Re e(\varepsilon) > 0$: strictly speaking, the surface response no longer originates in a "surface polariton", although it remains associated to an evanescent wave in vacuum.

In conclusion, we have shown that our "sharp resonance" approximation largely applies for various materials, and even when temperature broadening softens resonances. It yields a surface response simply estimated, over a limited range, by a complex Lorentzian derived from a linear extension of the optical constants. Notably, it implies a strict coupling between $\Re e S$ and $\Im m S$. Optical constants, as provided for discrete frequencies in [2], were also discussed in detail [4] for the evaluation of the surface resonances $\Re e S$ and $\Im m S$. The corresponding discrete plots, shown for AlSb, InSb, and YAG, featured dispersion-like and absorption-like resonances. They are now understood, on a firm basis, as originating in a single complex Lorentzian.

Further natural extensions of this very simple approach could concern the surface resonance at the interface between a dispersive material, and a transparent dielectric medium, with an index $n_2$: in this case, one has just to replace $\varepsilon(\omega)$ by $\varepsilon(\omega)/(n_2)^2$ in our reasoning. Also, a more complex geometry for the interface with vacuum can be considered. For a spherical interface, the surface response being governed by $(\varepsilon-1)/(\varepsilon+2)$, the approximate surface response is obtained for $\varepsilon(\omega) = -2$, and a sharp surface resonance approximately appears for $\kappa(\omega_{sphere}) = (2)^{1/2}$ and small values of $n$. A Lorentzian response, slightly shifted from $\omega_{sphere}$, is hence predicted in the same way as in eq. (5).

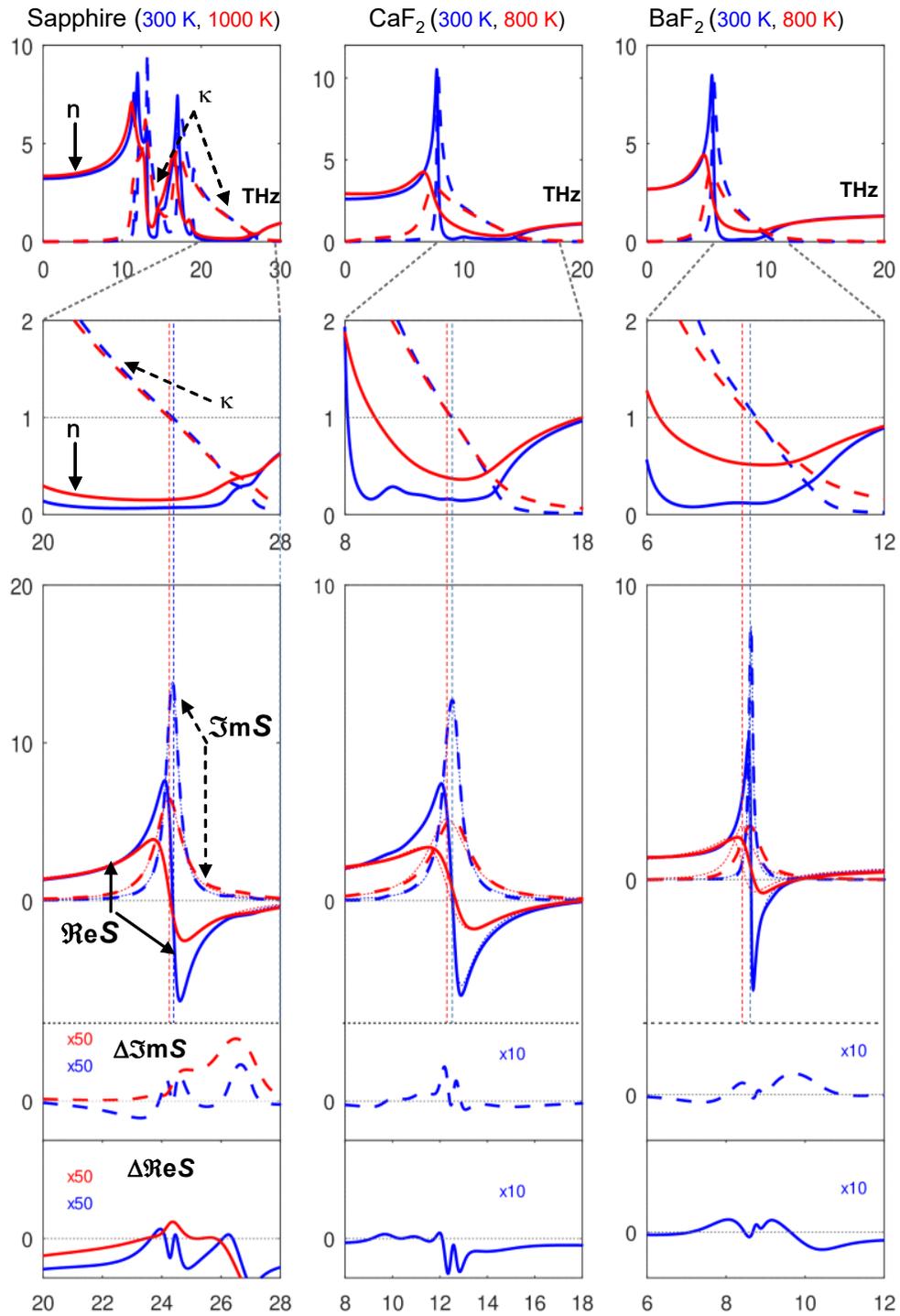

Fig. 1: Comparison between the surface response determined from literature and the Lorentzian response: from left to right, sapphire with perpendicular c-axis —as determined from [11] through $\varepsilon_{eff}=(\varepsilon_o\cdot\varepsilon_e)^{1/2}$ —, and $CaF_2$ and $BaF_2$ —both as determined from [9]. The top-line shows the $(n,\kappa)$ values (respectively full line, and dashed line) as a function of frequency, in blue (black when printed) for room temperature (~ 300 K), and in red (grey when printed) for high-temperature (~1000 K for sapphire, ~ 800 K for $CaF_2$ and $BaF_2$). The second line magnifies the first one, with scales restricted to a frequency range appropriate for the respective surface resonances. The faint vertical dashed lines (blue or red according to temperature) are markers for the frequencies satisfying $\kappa = 1 + n_0^2/2$ —i.e. $\omega = \omega_p - n_0^2/2K'$. On the last third line, $\Re S$ (full line) and $\Im S$ (dashed line) are shown as calculated from the $(n,\kappa)$ value. The faint dotted lines are the Lorentzian response calculated in our simplified approach. For sapphire (300 K and 1000 K), and for $CaF_2$ and $BaF_2$ at 300 K, the two calculations are hard to distinguish, and the magnified differences $\Delta\Im mS$ and $\Delta\Re S$, between the Lorentzian approximation and the complete calculation for $S$ are plotted —with a vertical shift, and a scale magnification.